\documentclass{Interspeech}

\interspeechcameraready

\title{Multi-interaction TTS toward professional recording reproduction}

\author[affiliation={},equalcontribution]{Hiroki}{Kanagawa}
\author[affiliation={},equalcontribution]{Kenichi}{Fujita}
\author[affiliation={}]{Aya}{Watanabe}
\author[affiliation={}]{Yusuke}{Ijima}

\affiliation[nocounter]{}{NTT, Inc.}{Japan}
\email{hiroki.kanagawa@ntt.com, kenichi.fujita@ntt.com}
\keywords{text-to-speech, expressive speech, speaking style refinement}

\usepackage{comment}
\usepackage{amsmath,graphicx,bm,amssymb,cite,url,multirow}
\usepackage{soul}
\usepackage[subrefformat=parens]{subcaption}
\usepackage{colortbl}
\usepackage{enumitem}

\begin{document}
\setlength\textfloatsep{7pt} 
\setlength\dbltextfloatsep{7pt} 
\setlength\floatsep{7pt} 
\setlength\abovecaptionskip{2pt} 
\setlength\belowcaptionskip{2pt} 
\captionsetup[subfloat]{aboveskip=2pt,belowskip=4pt} 

\maketitle

\begin{abstract}
Voice directors often iteratively refine voice actors' performances by providing feedback to achieve the desired outcome.
While this iterative feedback-based refinement process is important in actual recordings, it has been overlooked in text-to-speech synthesis~(TTS).
As a result, fine-grained style refinement after the initial synthesis is not possible, even though the synthesized speech often deviates from the user's intended style.
To address this issue, we propose a TTS method with multi-step interaction that allows users to intuitively and rapidly refine synthesized speech.
Our approach models the interaction between the TTS model and its user to emulate the relationship between voice actors and  voice directors.
Experiments show that the proposed model with its corresponding dataset enables iterative style refinements in accordance with users' directions, thus demonstrating its multi-interaction capability.

\end{abstract}

\section{Introduction}
\begin{figure}[tb]
  \centering
    \subfloat[Actual Recording]{%
    \includegraphics[width=0.44\linewidth]{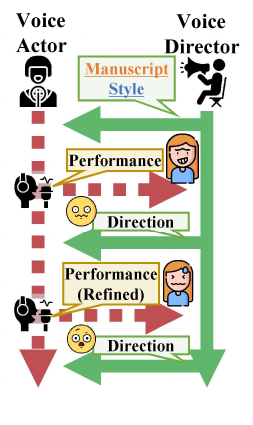}%
    \label{fig:recording}%
    }
  \hfill
  \subfloat[Proposed Method]{%
    \includegraphics[width=0.44\linewidth]{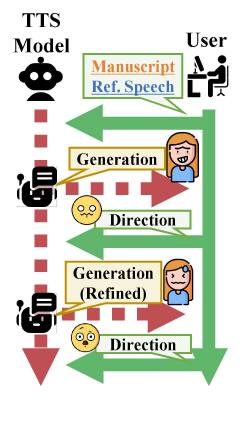}%
    \label{fig:proposed_method}%
  }
  \caption{Overview of the task and our proposed method.}
  \label{fig:overview_idea}
\end{figure}

Almost all creative activities entail a lot of trial and error to produce high-quality work~\cite{Tonnsen_2021,guss2021vinci,botella2018What}.
Film or television directors often require multiple takes, while photographers frequently adjust angles and camera settings.
Similarly, voice directors often iteratively refine voice actors' performance by providing feedback to achieve the desired speaking style~(see Fig.~\ref{fig:recording}) as in performance refinement of actors~\cite{glikpoe2023page,schmidt2023showing}.
Like the manual creation process, recent machine learning tasks now include iterative interactions between users and systems; examples include interactive image generation~\cite{ramesh2022dalle_2,rombach2022stable_diffusion,nijkamp2023text_to_image} and program coding~\cite{austin2021llm_programsynthesis,nijkamp2023codegen}.
This trend is being further accelerated by the emergence of large language models~(LLMs)~\cite{ouyang2022instruct_gpt,chowdhery2022palm}.
However, in speech synthesis, these iterative refinement processes have not been examined.

Recent text-to-speech synthesis~(TTS) has advanced rapidly, achieving human-level speech synthesis with naturalness comparable to that of recorded speech~\cite{tan2021tts_review,mehrish2023dnn_speech_review}.
In addition to naturalness, expressiveness has improved significantly, enabling the reproduction of diverse speaking styles.
Research into expressive TTS divides speech into three elements: text, speaker, and style information.
Starting from a simple style-ID based approach~\cite{INOUE202135} and the methods replacing pitch and durations from original speech~\cite{aylett19_ssw,vandevreken22_interspeech}, style information has recently been modeled using data-driven techniques with speech embedding~\cite{wang2018gst,lee2019finegrained,klimkov2019finegrained,zaidi2022daftexprt}.
Due to recent advances in LLMs, style information can be usefully inferred from text descriptions of speaking styles~\cite{Guo2023prompttts,Lyth2024synthetic_annotations,peng2024voicecraft}.
However, these methods do not support feedback from previously generated speech, requiring TTS users to iteratively select speech prompts or text prompts until the desired synthetic style is achieved.

To enhance the applicability of speech synthesis by enabling the easy generation of speech with the desired style, we believe incorporating iterative refinement is promising.
Building on this idea, we propose a new expressive TTS method guided by multiple directions~(see Fig.~\ref{fig:proposed_method}).
To incrementally refine TTS output towards the desired style, we treat the TTS model as a voice actor---akin to a voice actor converting text into speech---and design it to accept additional textual directions.
The speech synthesized using the provided direction is passed to the subsequent interaction.
Repeating this interaction efficiently, the style of the synthesized speech is refined.
To train our model, we designed a new dataset that replicates actual recordings.
This dataset includes multiple direction texts and voice actor recordings to simulate the process of style refinement.
These direction texts also includes not only manipulation of speaking style but also linguistic information (e.g., the placement of silent pause) unlike the similar text-prompted-TTS or voice conversion~(VC) approaches~\cite{Yan2025VoiceAttibute}.
During data collection, to minimize dependency on the voice director's skill and personality, the direction texts were prepared in advance rather than being given on the spot.

As a first step toward enabling human-like interactive iterative style refinement in expressive TTS, we conducted experiments focusing only on speech embedding manipulation guided by textual direction.
The results confirmed that our proposed method could approximately incorporate user's intended directions in the synthesized speech without degrading naturalness.
Moreover, our case analysis also highlighted remaining issues such as 1) specific position in a sentence and 2) dealing with ambiguous expressions, providing important insights for further style refinement improvement\footnote{
\url{https://ntt-hilab-gensp.github.io/ssw13multiinteractiontts/}}.

\section{Proposed method}
\begin{figure}[tb]
  \centering
  \includegraphics[width=\linewidth]{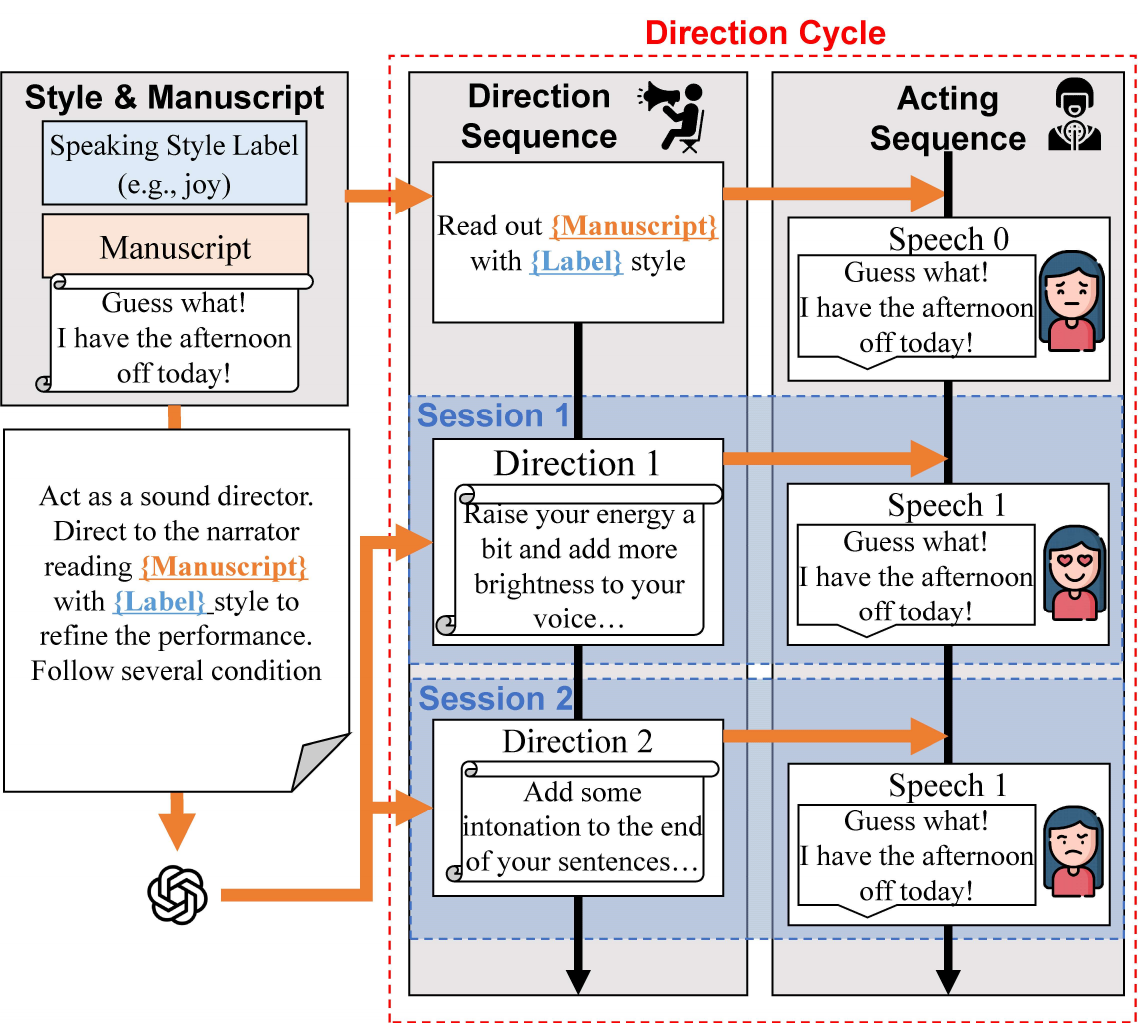}
  \caption{Overview of the direction cycle in dataset collection.}
  \label{fig:overview_dataset}
\end{figure}

\begin{figure}[tb]
  \centering
  \includegraphics[width=\linewidth]{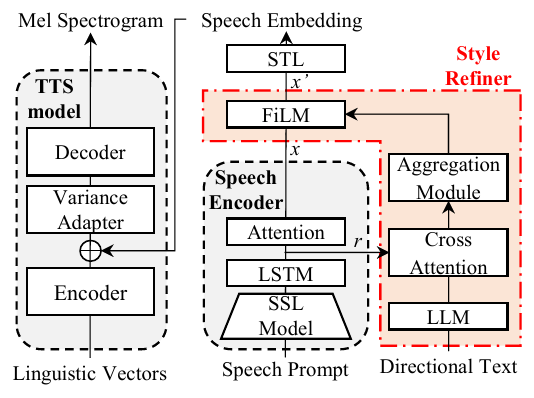}
  \caption{Overview of the proposed model.}
  \label{fig:overview_propose}
\end{figure}

To model the interaction between the voice director and the voice actor, we constructed a dataset that imitates an actual recording by them. 
During the recording of this dataset, a director iteratively gave acting directions, and the voice actor then reflected the given directions.
To model this direction cycle, we constructed a model capable of generating refined speech that reflects the given directions.
We would like to emphasize that style refinement is intended to include linguistic refinement in the future.
Therefore, our task is entirely different from voice conversion using text prompts~\cite{Yan2025VoiceAttibute}, in which only the speaker characteristics are changed while preserving the linguistic information.
This section explains the dataset~(Fig.~\ref{fig:overview_dataset}), backbone TTS model, and style refiner~(Fig.~\ref{fig:overview_propose}).

{\footnotesize
\begin{table}[tb]
  \caption{Example of style label, manuscript, and directions translated from Japanese.}
  \label{tab:example}
  \centering
  \resizebox{\linewidth}{!}{%
  \begin{tabular}{@{}c@{\hspace{5pt}} p{0.9\linewidth} }
    \toprule
      & Example\\
    \cline{1-2}
    Style Label &Resignedly \\
    Manuscript & I don't have time, okay? It's not like I wanted this either...\\\midrule\midrule
    Direction 1& Say it in a more restrained tone, with a sigh mixed in. Keep the delivery flat and steady, with little variation in strength. \\
    Direction 2& Let's try speaking so that your intonation gently falls at the end without sharp breaks, letting your words flow smoothly. Keep a slightly laid-back tone in mind.  \\
    Direction 3& Place just a slight pause between words, and speak in a tone that conveys resignation rather than urgency; overall, keep your pitch low.  \\
    \bottomrule
  \end{tabular}
  }
\end{table}
}
\subsection{Dataset}\label{sect:dataset}
To model interactions between the voice director and the voice actor, we constructed a dataset containing both speeches iteratively refined by the voice actor, and directions provided by the voice director~(Fig.~\ref{fig:overview_dataset}).

In actual voice recording sessions, voice directors provide acting instruction to voice actors in various ways. Such instruction can be categorized as follows:
\begin{enumerate}
    \item Instructions for modifying para-linguistic and non-linguistic information in the acted speech~(e.g., ``Speak more brightly,'' ``Be more considerate of the other person's feelings'').
    \item Instructions for modifying linguistic information~(e.g., ``Insert a silent-pause after this word,'' ``Change the location of stress~(or accent~(for pitch-accent language, i.e., Japanese))'').
    \item Demonstrative instructions, where the director provides a performance example~(e.g., ``Follow my example'').
    \item Instructions involving gestures and body language.
\end{enumerate}
While it would be ideal to construct a dataset that contains all of these categories, creating such a comprehensive dataset is highly complex and presents substantial challenges. Therefore, in this study, we constructed a dataset limited to categories 1 and 2, which can be expressed solely in textual form.

Furthermore, in most actual voice recording sessions, voice directors iteratively review the results and provide instructions to guide voice actors toward the desired outcome. However, replicating this process in a dataset is difficult, as the diversity and consistency of instruction vocabulary heavily depend on the director's expertise.
Thus, this approach would be unsuitable for building a large-scale dataset.

To reliably collect a large amount of data without relying on the skill of the voice director, we generated the directions using an LLM without referring to the actual recordings.
This reduces the effect of the voice director's personal style and makes it easier to obtain a large dataset while maintaining direction quality.

To construct the dataset including various speaking styles, we prepared manuscripts aligned with style categories~(e.g., sleepy, child-oriented, and urgent) following ``The Guideline for TTS Speaking Style Classification''~\cite{JEITA}.
Based on these speaking styles, we generated directions for each manuscript by entering text prompts for an LLM~(examples are available on our demo page).
The LLM is instructed to act as a voice director and generate iterative directions two or three times.
For quality validation, three native Japanese speakers manually verified the formatting and overall quality.
Different from text prompting TTS methods such as PromptTTS~\cite{Guo2023prompttts}, our generated directions include more complex and abstract text descriptions; the text prompts in past research are limited to simple directions e.g., high/low pitch, fast/slow speech rate.
Moreover, the directions cover the instructions for changing the linguistic information, e.g., inserting pauses. 

\subsection{Backbone TTS model}
\label{sect:ssl_model}
The TTS model is conditioned on an embedding extracted from a target speech. 
To obtain this embedding, we employed a speech encoder and a style token layer~(STL)~\cite{wang2018gst}. 
Specifically, the speech encoder leverages a self-supervised learning~(SSL) speech model~\cite{fujita2023zeroshot} to extract feature representations, which are then aggregated into a single vector, denoted as $\boldsymbol{x}$, using a weighted sum~\cite{chen22g_interspeech}, a bidirectional LSTM, and an attention mechanism~\cite{raffel2016feedforward}.
Subsequently, STL post-process $\boldsymbol{x}$ was applied to enhance stability, resulting in the final speech embedding. By jointly training the TTS model with these modules, the speech embedding so generated becomes well-suited for TTS systems.

\subsection{Style refiner}\label{sect:style_updater}
The style refiner refines $\boldsymbol{x}$, the pre-refined embedding, into $\boldsymbol{x'}$, the refined embedding, using information from the direction text.
The style refiner consists of three components: cross-attention, aggregation module, and FiLM~\cite{FiLM_2018}.
Cross attention integrates information from speech and text, as in other studies on text-speech integration~\cite{xu19c_interspeech,9414654}.
To handle richer speech information, cross attention takes as input
$\boldsymbol{r} \in {\mathbb R}^{D_r \times L_r }$ , output from an intermediate LSTM layer in the speech encoder instead of $\boldsymbol{x} \in {\mathbb R}^{D_x \times 1}$.
Here, $D_r$, $L_r$, and $D_x$ represent the dimension and number of frames of $\boldsymbol{r}$, and the dimension of $\boldsymbol{x}$, respectively.
After integration, the aggregation module, which uses the same attention mechanism as in the speech encoder, aggregates the representations into a single fixed-dimensional embedding.
Finally, $\boldsymbol{x}$ is converted into $\boldsymbol{x'}$ via a FiLM layer conditioned on the aggregated representation.

The \textit{style refiner} is trained separately, with the backbone TTS model trained first.
We extracted $\boldsymbol{x}$ and $\boldsymbol{r}$ from the pre-refined speech and $\boldsymbol{x'}$ from the refined speech.
These extracted representations and the corresponding direction text are paired and used to train the style refiner.
The L1 loss between the target and predicted $\boldsymbol{x'}$ is utilized.
At inference time, the \textit{style refiner} produces $\boldsymbol{x'}$, which is then passed to STL and converted into a speech embedding.
This embedding is then fed into the TTS model to generate speech that is expected to be the modified version of the pre-refined speech refined according to the given direction.

\section{Experimental setup}
\subsection{Dataset}
We used three Japanese \SI{22}{\kHz} datasets: interactive, non-interactive, and large in-house datasets.
The interactive and non-interactive datasets are from the same two voice actors (one female and one male).
The interactive dataset was specially designed for this study as described in Sect.~\ref{sect:dataset}. It includes 13.7 hours of Japanese speech, consisting of 5,639 utterances collected from 1,606 direction cycles.
From this dataset, 104 direction cycles~(all 52 styles $\times$ 2 voice actors) were selected for validation and another 104 for testing. 
The direction texts were prepared by using ChatGPT-4o~\cite{gpt4o} and the style labels were selected from 52 speaking styles~(neutral, screaming, and laughing were excluded from the original 55 speaking styles in the guideline~\cite{JEITA}). During the recording, for the real imitation, the future directions were concealed from the voice actors, and the necessary directions were provided to them one-by-one.

The non-interactive dataset is the standard TTS dataset without interactive sessions, including 43.6 hours of speech data consisting of 42,039 utterances.
The dataset includes acted speech in the 55 speaking styles following the guideline~\cite{JEITA}.
The large in-house dataset includes 8K-hours of Japanese speech. 

To improve the style refiner's robustness to the diversity of the directions, we adopted four LLM-based augmentation methods. 
The first one is \textit{Simple}; it which summarizes the original directions by LLM, assuming that careful directions cannot be obtained from users.
The other three were through prompts proposed by Maini et al. i.e., using their prompt templates; \textit{Easy}, \textit{Hard}, and \textit{Medium Style}~(excluding \textit{Q/A Style})~\cite{maini2024rephrasing}.
As the dataset is in Japanese, we used the translated versions of these templates.
We used Llama-3.1-Swallow-8B-Instruct v0.2\footnote{\url{https://huggingface.co/tokyotech-llm/Llama-3.1-Swallow-8B-Instruct-v0.2}} as the LLM, and the augmented directions eventually yielded seven per original direction~(Two with \textit{Easy}, \textit{Hard}, and \textit{Medium}, and one with \textit{Simple} prompt).

\subsection{Training conditions}
The TTS model was based on FastSpeech2~\cite{ren2020fastspeech}, following the implementation described in a previous work~\cite{chien2021investigating}.
The input of the TTS model was 303-dimensional linguistic vectors with phoneme and accent information.
The output was an 80-dimensional log-mel spectrogram with \SI{10}{ms} frame shift.
We incorporated a HuBERT \textsc{Base} model, pre-trained on ReazonSpeech\footnote{\url{https://huggingface.co/rinna/japanese-hubert-base}}, with its parameters frozen.
This model converted each \SI{16}{kHz} raw audio sample into a sequence of 768-dimensional vectors, which were then transformed into fixed-length 384-dimensional \textit{speech embeddings}. 
The speech waveform was generated by HiFi-GAN~(V1)~\cite{NEURIPS2020_c5d73680}.
As the style refiner, we utilized Gemma2 LLM fine-tuned on a Japanese dataset\footnote{\url{https://huggingface.co/rinna/gemma-2-baku-2b-it}}. 
Different LLMs were used for the style refiner (Gemma2), direction text generation (ChatGPT-4o) and augmentation (Llama-3.1-Swallow-8B-Instruct v0.2) to prevent performance collapse induced by training on biased knowledge from a single model, as reported in \cite{shumailov2024curserecursiontraininggenerated}.

The proposed model was trained in four steps as follows.
\begin{enumerate}
  \item The backbone TTS model was trained in 600K steps using the large in-house dataset.
  \item The backbone TTS model was further trained with GAN in the same manner as~\cite{kanagawa19_ssw,peng2018variational} for an additional 200K steps with a fixed learning rate of $1 \times 10^{-4}$ to improve the naturalness of the synthesized speech. 
  \item The model was then fine-tuned in 30K steps with the interactive and the non-interactive datasets to improve the speaking style reproduction for each voice actor. The loss function and learning rate were the same as in the previous step.
  \item The style refiner was trained by the interactive dataset. The training step was 10K steps with AdamW optimizer~\cite{loshchilov2018decoupled} with 4K warm-up steps.
\end{enumerate}

\begin{table*}[t]
    \caption{
    Speech generation conditions in a direction cycle across three sessions, indicating how different methods (Identical, Actor-Guided, Single-shot, and Iterative) utilize speech prompts and directions.}
    \setlength{\tabcolsep}{4pt}
    \label{table:gen_cond}
    \centering
    \resizebox{\linewidth}{!}{%
    \begin{tabular}{@{}l@{\hspace{2pt}}c*{9}{c}@{}}
        \toprule
        Conditions&\multicolumn{3}{c}{Session~1}&\multicolumn{3}{c}{Session~2}&\multicolumn{3}{c}{Session~3}\\\cmidrule(l{\tabcolsep}){2-4}\cmidrule(l{\tabcolsep}){5-7}\cmidrule(l{\tabcolsep}){8-10}
        &\multicolumn{2}{c}{Input}&Output&\multicolumn{2}{c}{Input}&Output& \multicolumn{2}{c}{Input}&Output&\\
        &Speech Prompt&Direction~1&Speech~1&Speech Prompt&Direction~2&Speech~2&Speech Prompt&Direction~3&Speech~3&\\
        \midrule\midrule
        Identical&\cellcolor[gray]{0.9} Recorded-0&-&\textbf{Identical-0}&\cellcolor[gray]{0.9}Recorded-0&-&\textbf{Identical-0}&\cellcolor[gray]{0.9}Recorded-0&-&\textbf{Identical-0}&\\
        Actor-Guided~(oracle)~~&\cellcolor[gray]{0.9}Recorded-1&-&\cellcolor[gray]{0.8} Guided-1&\cellcolor[gray]{0.9}Recorded-2&-&\cellcolor[gray]{0.8}Guided-2&\cellcolor[gray]{0.9}Recorded-3&-&\cellcolor[gray]{0.8}Guided-3&\\
        \midrule
        Single-shot~(ours)&\cellcolor[gray]{0.9}Recorded-0&\checkmark&Single-1&\cellcolor[gray]{0.9}Recorded-1&\checkmark&Single-2&\cellcolor[gray]{0.9}Recorded-2&\checkmark&Single-3&\\
        Iterative~(ours)&\cellcolor[gray]{0.9}Recorded-0&\checkmark&\textit{Iterative-1}&\textit{Iterative-1}&\checkmark&\textit{Iterative-2}&\textit{Iterative-2}&\checkmark&\textit{Iterative-3}&\\
        \bottomrule
    \end{tabular}
    }
\end{table*}

\section{Subjective evaluation}
We conducted three subjective evaluations to confirm the effectiveness of the proposed method. 
The evaluations examined iterative style refinement, refinement accuracy, and naturalness.
For the evaluation, we split the test dataset~(52 speaking styles) into five style groups as reported in~\cite{homma2023expressive}.
These groups can be interpreted as groups of fear, joy, anger, sadness, and surprise.

\subsection{Speech preparation}
The speech samples were generated under four conditions: \textit{Identical}, \textit{Actor-Guided~(oracle)}, \textit{Single-shot~(ours)}, and \textit{Iterative~(ours)} in Table~\ref{table:gen_cond}.
During speech generation, we used two types of speech prompts: actually-recorded speech~(\textit{Recorded-N}) and generated speech~(\textit{Iterative-N}, \textit{Identical-0} and \textit{Guided-N}), where \textit{N} indicates that the speech was taken from the \textit{N}-th session of the direction cycle.

In both the \textit{Iterative~(ours)} and \textit{Single-shot~(ours)} conditions, our style refiner received speech prompts (iteratively or non-iteratively generated in prior sessions) and the corresponding direction text.
In the \textit{Identical} and \textit{Actor-Guided~(oracle)} conditions, speech was generated without using our style refiner~(i.e., normal TTS generation), using actually recorded speech prompts.

\subsection{Iterative style refinement}\label{sect:test_iteratice_dmos}
We conducted a subjective evaluation of iterative style refinement.
Participants saw a pseudo-direction-cycle of up to three refinement iterations under three conditions: \textit{Identical}, \textit{Actor-Guided~(ours)}, and \textit{Iterative~(ours)} in Table~\ref{table:gen_cond}.
Each pseudo-direction cycle corresponds to a cycle in the test data~(Fig.~\ref{fig:overview_dataset}).
Participants saw the direction text, then \textit{reference speech} and \textit{test speech}, forming a pseudo-direction cycle~(see Table~\ref{table:dmos_given_audio} for details).
Participants then rated each pair on a five-point MOS test with the direction text being used: 1 - No change; 2 - Minor modification; 3 - Overall alignment only; 4 - Adjustment reflecting the director's intent; 5 - Significant change fully embodying the director's intent.
Prior to the evaluation, participants were given concrete examples corresponding to each score.
For instance, a score of 3 (``Overall alignment only'') was explained as a case where the speech generally matched the emotional tone or style requested, but lacked specific refinements.
The \textit{Identical} condition served to validate our evaluation baseline  by rating identical speech pairs.
We expect the directions to be correctly reflected in actual recordings, so the \textit{Actor-Guided~(oracle)} condition represents the upper bound of our method.

\begin{table}[t]
    \caption{The paired speech from a direction cycle from the evaluations in ~Sect.~\ref{sect:test_iteratice_dmos} and \ref{sect:test_cmos}.}
    \label{table:dmos_given_audio}
    \setlength{\tabcolsep}{5pt}
    \centering
    \resizebox{\linewidth}{!}{%
        \begin{tabular}{@{}l@{\hspace{4pt}}c*{5}{c}@{}}
        \midrule
        \multicolumn{2}{c}{Method} & Identical & Actor-guided &Single-shot&Iterative \\ \hline
        \multicolumn{2}{c}{Style refiner} &- & -&\checkmark&$\checkmark$ \\ 
        \midrule\midrule
        Session~1 & Ref. Speech & \textbf{Identical-0}& \textbf{Identical-0} &\textbf{Identical-0}& \textbf{Identical-0} \\
        &{Test Speech} & \textbf{Identical-0} & \cellcolor[gray]{0.8} Guided-1 &Single-1&\textit{Iterative-1} \\ 
        \midrule
        Session~2& Ref. Speech & \textbf{Identical-0} &\cellcolor[gray]{0.8} Guided-1 &\cellcolor[gray]{0.8}Guided-1& \textit{Iterative-1} \\
        &{Test Speech} & \textbf{Identical-0} & \cellcolor[gray]{0.8}Guided-2 &Single-2& \textit{Iterative-2} \\ 
        \midrule
        Session~3& Ref. Speech & \textbf{Identical-0} &\cellcolor[gray]{0.8} Guided-2 &\cellcolor[gray]{0.8}Guided-2&\textit{ Iterative-2} \\
        & {Test Speech} & \textbf{Identical-0} & \cellcolor[gray]{0.8}Guided-3 &Single-3&\textit{Iterative-3} \\ 
\bottomrule
\end{tabular}
}
\end{table}


\begin{figure}[!t]
  \centering
  \includegraphics[width=\linewidth]{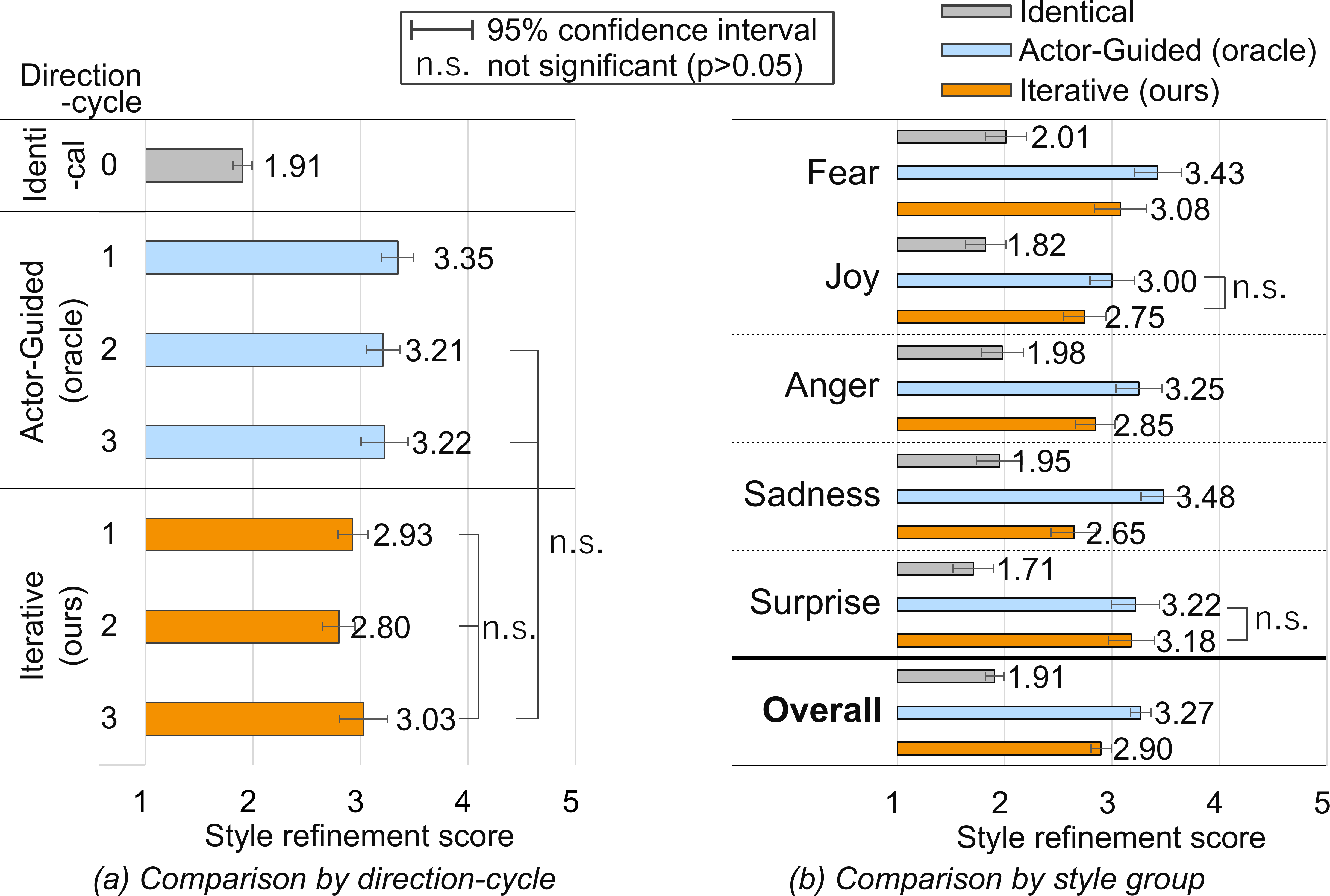}
  \caption{Subjective evaluation results for iterative style refinement. (a) compares each method regarding direction-cycle. (b) breaks them down into five style groups and shows the overall results.}
  \label{fig:dmos_style_refinement}
\end{figure}
The evaluation was conducted with 82 participants via a crowdsource interface.
For the evaluation, we selected 16 direction cycles for each method, i.e., three or four cycles per subgroup from each voice actor of the test dataset.
The results are shown in Fig.~\ref{fig:dmos_style_refinement}~(a).
\textit{Iterative~(ours)} significantly outperformed \textit{Identical} due to the effectiveness of our style refiner.
On the other hand, \textit{Iterative~(ours)} slightly underperformed \textit{Actor-Guided~(oracle)}.
These results imply that our proposed method achieves style refinement that aligns with the intended refinement.
The gap between \textit{Actor-Guided~(oracle)} and \textit{Iterative~(ours)} implies that there is still a discrepancy from the desired style.
The \textit{Iterative~(ours)} condition also showed no significant score change across direction cycles, implying that directions are reflected in the synthesized speech regardless of the number of previous refinements.

Figure~\ref{fig:dmos_style_refinement}(b) breaks down the aggregated results into style groups, and also shows the overall results.
\textit{Iterative~(ours)} scored lower than \textit{Actor-Guided~(oracle)} in all five style groups.
While no significant difference was found between both methods in ``Joy'' and ``Surprise'' style groups, the difference between them was significant in ``Overall'' as well as the other style groups.
These results suggest that \textit{Iterative~(ours)} still has room for improvement in terms of style refinement, as indicated earlier in Fig.~\ref{fig:dmos_style_refinement}(a).

\subsection{Style refinement accuracy}\label{sect:test_cmos}
We then evaluated whether the refinement truly aligned with the given directions.
Participants were presented textual direction, and then continuously presented pre-refined and refined speeches.
The participants then evaluate the difference in speech alignment with the given text by a seven-point MOS test ranging from \textminus3~(refinement deviates from the direction) to 3~(refinement aligns with the direction).
The speech was generated under \textit{Single-shot~(ours)} condition in Table~\ref{table:gen_cond}, and pre-refined and refined speeches corresponding to \textit{reference} and \textit{test speech} of \textit{Single-shot~(ours)} condition in Table~\ref{table:dmos_given_audio}.

We prepared two conditions: \textit{Matched} and \textit{Random}.
In the former condition, the direction presented to participants was actually used for refinement, and in the latter, the presented direction was randomly selected from the test data.
We selected 50 sessions for each voice actor (10 sessions for each of five subgroups).
Each session was evaluated three times: once under the \textit{Matched} condition and twice under the \textit{Random} condition.
To cover sentence diversity, \textit{Random} condition was evaluated with two distinct directions.
The number of participants was 110 gathered through the same crowdsourcing interface.
Moreover, for detailed analysis, we classified the \textit{Random} condition into \textit{Random~(Similar)} and \textit{Random~(Dissimilar)} using ChatGPT o3-mini.
We asked ChatGPT o3-mini to score the similarity, ranging from 1~(dissimilar) to 5~(similar), between the original direction used for refinement and the one randomly selected.
From these predicted scores, we regard sentences that scored two or lower as \textit{Random~(Dissimilar)} and those scoring three or higher as \textit{Random~(Similar)}.


\begin{figure}[!t]
  \centering
  \includegraphics[width=0.8\columnwidth]{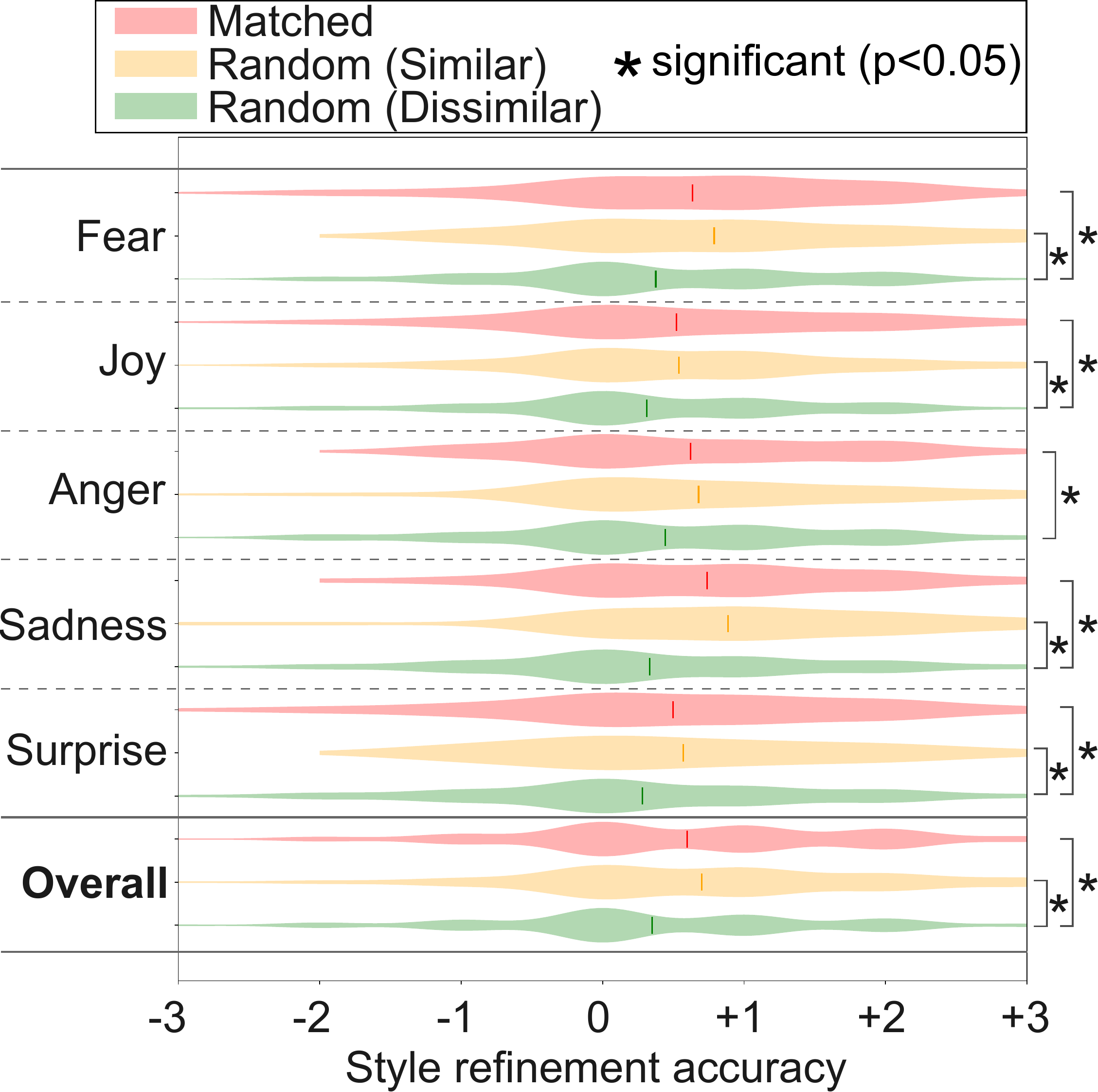}
  \caption{Subjective evaluation results on style refinement accuracy. The ``Overall'' result was obtained from all evaluation scores, without style group categorization.}
  \label{fig:cmos_style_refinement}
\end{figure}

Figure~\ref{fig:cmos_style_refinement} shows a violin plot of style refinement accuracies for each style and each condition using the raw scores obtained from the participants.
\textit{Matched} significantly outperformed {\textit{Random~(Dissimilar)} for all style groups.
\textit{Random~(Similar)} yield comparable performance to {\textit{Matched}}.
These results reveal that even if the direction did not exactly match between that used for style refinement and that given to listeners, comparable performance could be obtained if their directions were similar.
On the other hand, when the direction sentence given to listeners was completely different from that  used for style refinement~(i.e., \textit{Random~(Dissimilar)}), the score was significantly worse than that of \textit{Matched} and \textit{Random~(Similar)}, and the score distribution was skewed lower.
These results show that the refinement achieved by our method is consistent with the given directions.

\subsection{Naturalness}\label{sect:naturalness}
We finally analyzed the effect of iterative generation on naturalness using a MOS on a five-point scale~(1: very unnatural, 5: very natural).
We compared three conditions: \textit{Single-shot~(ours)}, \textit{Iterative~(ours)}, and \textit{Actor-Guided~(oracle)} in Table~\ref{table:gen_cond}.
The number of participants was 168 via the same crowdsourcing interface. We selected 15 sentences for each condition i.e., three sentences per subgroup of the test dataset from each voice actor. 

\begin{figure}[t]
  \centering
  \includegraphics[width=\linewidth]{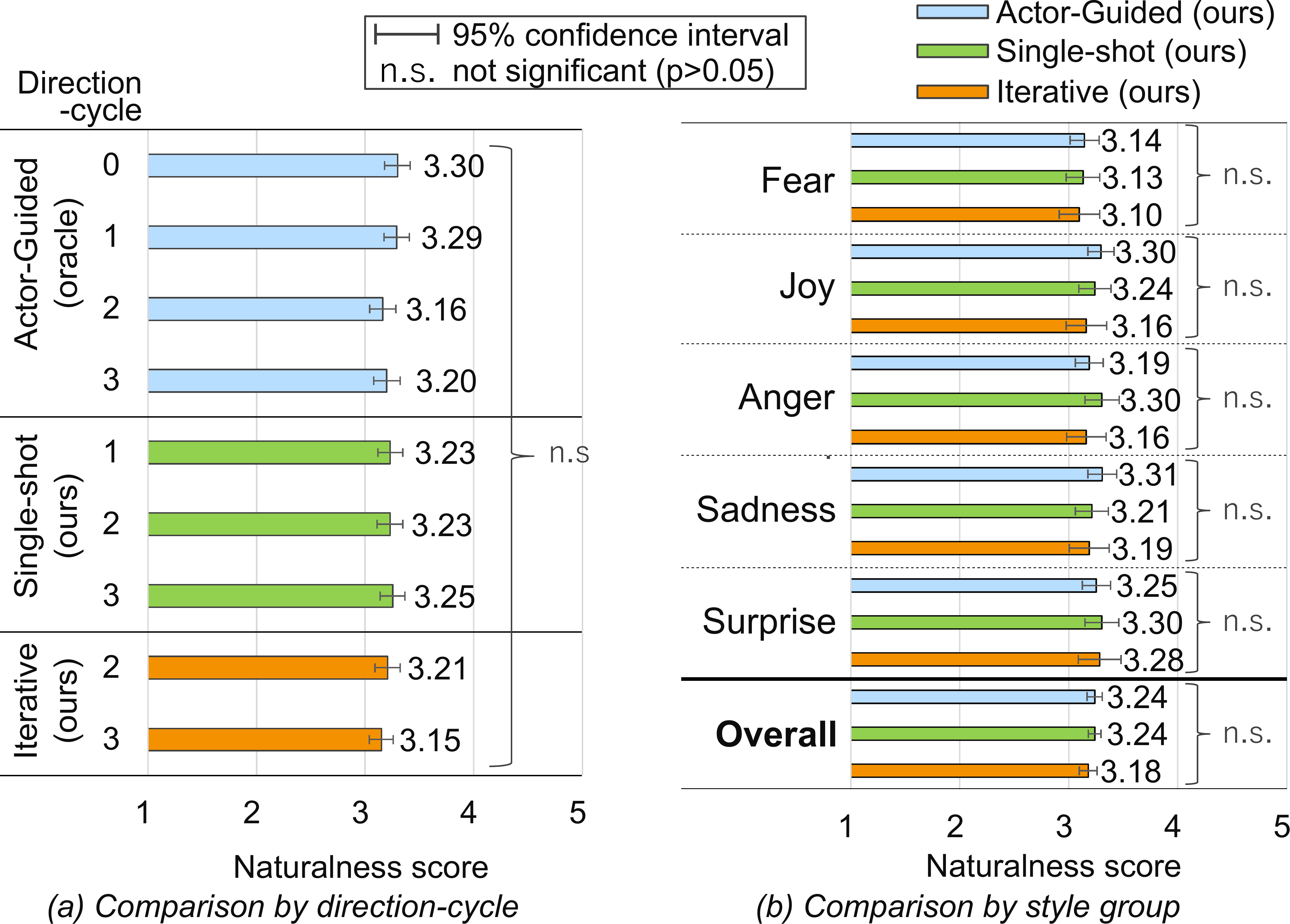}
  \caption{Subjective evaluation results for naturalness. Similar to Fig.~\ref{fig:dmos_style_refinement}, (a) compares all methods over direction-cycle. (b) shows both the results for style group and ``Overall'' obtained from all evaluation data, without style group categorization.}
  \label{fig:mos_style_refinement}
\end{figure}
The results are shown in Fig.~\ref{fig:mos_style_refinement}. 
All conditions, i.e., \textit{Actor-Guided~(oracle)}, \textit{Single-shot~(ours)}, and \textit{Iterative~(ours)}, show no significant difference in terms of naturalness.
This implies that iterative refinement explicit in our method does not degrade naturalness.

{\footnotesize
\begin{table}[tb]
  \caption{The directions for top and worst three samples in the interactive style refinement evaluation of \textit{Interactive~(ours)}. \ul{under-lined} and \textbf{bold} parts represent, respectively, \ul{instruction for the specific words or positions} and \textbf{instructions for changing linguistic information}. Scores are from Sect.~\ref{sect:test_iteratice_dmos}.}
  \label{tab:difficult}
  \centering
  \resizebox{\linewidth}{!}{%
  \begin{tabular}{@{}c@{\hspace{5pt}} p{0.9\linewidth} }
    \toprule
    Score & Direction~(Translated from Japanese) \\
    \cline{1-2}
    4.67 &Try speaking bit faster overall. Especially in \ul{the latter half}, picking up the pace slightly will help emphasize the joy of discovery \\\cmidrule(l{\tabcolsep}){1-2}
    4.22 & \ul{At the beginning of the line}, try \textbf{taking a small breath in}, as if you're slightly out of breath. This will help convey a sense of excitement.\\\cmidrule(l{\tabcolsep}){1-2}
    4.20  &Slow down the speaking pace a little, and try to let the sound fade out \ul{at the end}, as if it's gently disappearing.  \\\midrule\midrule
    2.30& \textbf{Hold your breath} slightly \ul{at the end}, as if swallowing your words. Make it sound more desperate and urgent.  \\\cmidrule(l{\tabcolsep}){1-2}
    2.30&Lower your tone \ul{at the end} and let your voice fade out weakly. This will create a worn-out impression.  \\\cmidrule(l{\tabcolsep}){1-2}
    2.08 &Extend the ending slightly to add a gentle, inviting nuance. \ul{For the part of ``how about doing''}, speak softly as if warmly encouraging the listener.  \\
    \bottomrule
  \end{tabular}
  }
\end{table}
}

\section{Discussion}
As described in Sect.~\ref{sect:dataset}, our dataset includes difficult directions that were not included in previous text-prompt-based TTS methods.
We believe that this explains why the overall score in Sect.~\ref{sect:test_iteratice_dmos} is relatively low; it remains around three even in non-iterative settings~(\textit{Actor-Guided~(oracle)}).

Table~\ref{tab:difficult} presents examples from the test data that received the top and bottom three scores in the experiment in Sect.~\ref{sect:test_iteratice_dmos}.
This section addresses the issues and suggests solutions for future work by examining specific examples in the dataset.

\subsection{Style modeling}
This study primarily focused on style refinement through iterative sessions.
However, we found that some examples, especially those with low scores, include instructions that are hard to achieve with a single speech embedding-based TTS system.
Specifically, some include instructions that designate specific words or positions in the manuscript~(e.g., ``at the beginning'', ``at the end'', and ''for the part of `how about doing''').
Since global embeddings generally cannot alter the expression of specific words within a sentence, such instructions are difficult to reflect through refinement with our method.
One promising solution would be to use a fine-grained approach, such as NaturalSpeech~3~\cite{ju2024naturalspeech} or VALL-E~\cite{wang2023neural}, for the TTS model.

\subsection{Linguistic contents modeling}
When giving instructions for acting, support for changing the linguistic information is also important.
In fact, some directions include instructions for these features~(e.g., ``hold your breath'' and ``place just a slight pause between words'').
However, our current model refines just the style of the speech while maintaining linguistic information.
To achieve this support, it is desirable to refine not only paralinguistic information but also linguistic information from instructions. One promising approach would be to incorporate LLM for linguistic information modification into our proposed method.

\subsection{Design of subjective evaluation}
The evaluations in this paper examine overall impressions.
This is reasonable given that this is the initial evaluation of our proposal.
However, as this evaluation method may not fully capture subtle stylistic differences, the relatively low scores from the \textit{Actor-Guided~(oracle)} condition in Sect.~\ref{sect:test_iteratice_dmos} is reasonable.
Recent text-to-image/video generation tasks also face the same evaluation difficulties.
Approaches for capturing detailed production differences to assess alignment with the text prompt~\cite{NEURIPS2023_c481049f,Otani_2023_CVPR} are being studied.
Similarly, further exploration of more fine-grained and appropriate evaluation methods are required to advance our work.

\section{Conclusion}
In this paper, we proposed a method for TTS systems that can model the interaction between voice directors and voice actors. Subjective evaluations demonstrated that our proposed method achieved iterative style refinement that matched the user's directions to some extent.
Moreover, analysis of several experiments gave suggestions on further enhancement of our proposed method toward achieving human-level style refinement. 
Our future work will involve further refinement of style, not only through speech embedding manipulation but also by incorporating linguistic information (e.g., the placement of silent pauses, stress, or pitch accents). Investigating more effective evaluation methods for this task remains an ongoing challenge. As an initial step towards a style refinement framework, we examined a speaker-dependent style refiner, and we plan to extend this approach to the speaker-independent scenario.

\bibliographystyle{IEEEtran}
\bibliography{mybib,refs_kf}
\end{document}